\documentclass[prl,amssymb,twocolumn,tightenlines,showpacs]{revtex4-1}

\usepackage{graphicx}
\usepackage{amsmath}
\usepackage{verbatim}
\usepackage{subfigure}

\begin{document}

\title{A Resource Efficient Source of Multi-photon Polarization Entanglement}

\author{E. Megidish}
\affiliation{Racah Institute of Physics, Hebrew University of
Jerusalem, Jerusalem 91904, Israel}
\author{T. Shacham}
\affiliation{Racah Institute of Physics, Hebrew University of
Jerusalem, Jerusalem 91904, Israel}
\author{A. Halevy}
\affiliation{Racah Institute of Physics, Hebrew University of
Jerusalem, Jerusalem 91904, Israel}
\author{L. Dovrat}
\affiliation{Racah Institute of Physics, Hebrew University of
Jerusalem, Jerusalem 91904, Israel}
\author{H. S. Eisenberg}
\affiliation{Racah Institute of Physics, Hebrew University of
Jerusalem, Jerusalem 91904, Israel}


\begin{abstract}
Current photon entangling schemes require resources that grow with
the photon number. We present a new approach that generates
quantum entanglement between many photons, using only a single
source of entangled photon pairs. The different spatial modes, one
for each photon as required by other schemes, are replaced by
different time slots of only two spatial modes. States of any
number of photons are generated with the same setup, solving the
scalability problem caused by the previous need for extra
resources. Consequently, entangled photon states of larger numbers
than before are practically realizable.
\end{abstract}

\maketitle

The generation of quantum entangled states of many particles is a
central goal of quantum information science. These states are
required for the one-way quantum computer scheme
\cite{Raussendorf01,Walther05}. In quantum communication, they
enable error correction \cite{Schlingemann02} and multi-party
protocols \cite{Hillery99,Zhao04}. Additionally, they can refute
local realistic theories with an increasing violation as the
particle number increases \cite{Mermin90,Zukowski97,Guhne05}.

Polarized photons are an attractive realization of qubits due to
their simple single-qubit operations and their weak interaction
with the environment. Pairs of polarization entangled photons are
easily generated by using the nonlinear optical effect of
parametric down-conversion (PDC) \cite{Kwiat95}. Difficulties are
encountered when trying to entangle more than two photons. Eight
photons have already been entangled in a GHZ state
\cite{Greenberger90,Huang11,Yao12}, and six photons in an H-shaped
graph \cite{Lu07} and Dicke \cite{Wieczorek09} states. Currently,
the state with the largest number of entangled particles of any
realization is a GHZ state composed of 14 ions trapped in a linear
trap \cite{Monz11}.

\begin{figure*}
\centering\includegraphics[angle=0,width=\textwidth]{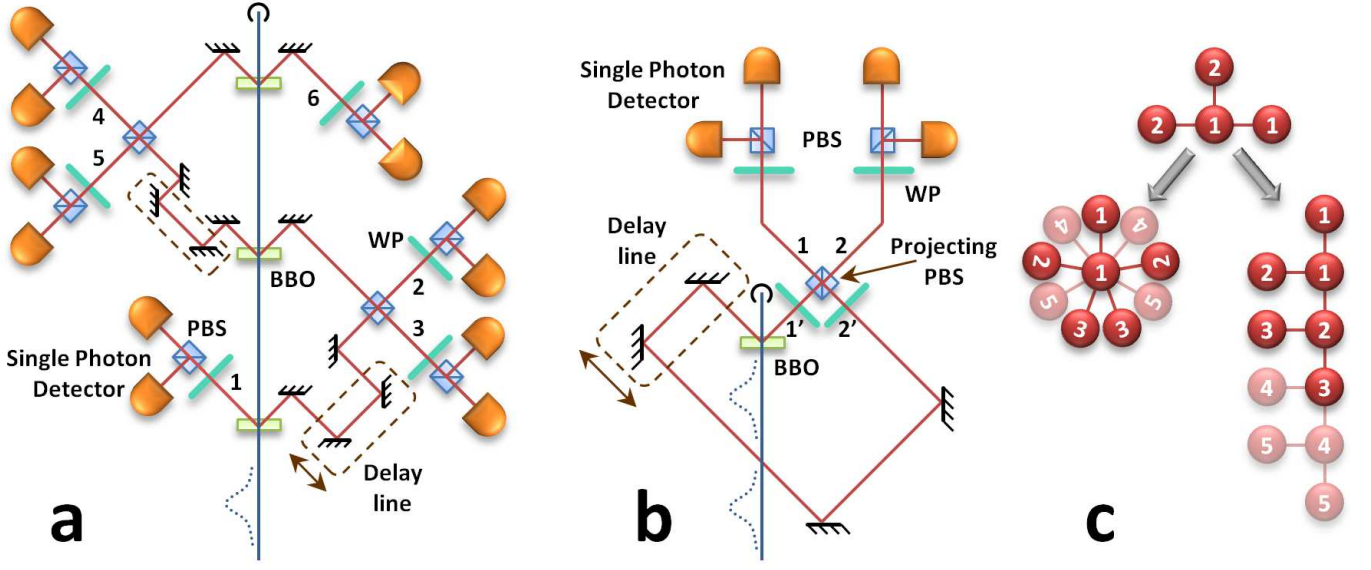}
\caption{\label{Fig1}(Color online) A comparison between previous
setups and our setup. (a) The previously used setup for entangling
four and six photons in a GHZ state. Three $\beta$-BaB$_2$O$_4$
(BBO) crystals generate from a single pump pulse three entangled
photon pairs in six spatial modes. Delay lines synchronize the
projection of a photon from each pair onto a PBS. Six analyzing
wave-plates (WP), six PBS's and twelve detectors are required to
measure the state. To produce eight photon entanglement, this
setup needs additional components. (b) Our resource efficient
setup. Only a single crystal generates pairs from many pump
pulses. The pairs are projected onto a large entangled state on a
single PBS and occupy two spatial modes and additional temporal
modes. The same setup is applicable to entanglement generation of
any photon number. (c) The two multi-photon entangled graph states
that are possible to obtain without fast polarization rotations: a
growing star shaped GHZ state and a connected branched chain
(which for 3 pairs is an H-graph state). For two photon pairs,
both states are identical. Numbered circles mark photons and their
creation order, where dimmed circles represent possible future
states of 4 and 5 fused pairs. Connecting lines mark the
entangling operations that define the quantum graph state.}
\end{figure*}

In previous experiments, two polarization-entangled photon pairs
were fused into a four-photon GHZ state by a polarizing
beam-splitter (PBS) \cite{Pan01} (see Fig. \ref{Fig1}a). A PBS is
an optical element that transmits horizontally ($h$) polarized
photons and reflects them when vertically polarized ($v$).
Assuming an entangled pair in the $|\phi^+\rangle$ Bell state
\cite{Kwiat95} in paths 1 and 2, and another one in paths 3 and 4,
the four-photon state is:
\begin{equation}\label{product}
|\phi_{12}^+\rangle\otimes|\phi_{34}^+\rangle=\frac{1}{2}(|h_1h_2\rangle+|v_1v_2\rangle)\otimes(|h_3h_4\rangle+|v_3v_4\rangle)\,.
\end{equation}
If paths 2 and 3 are combined at a PBS, and we demand that one
photon comes out of each of the two output ports, only the two
amplitudes with identical polarizations in these modes are left
and the result is a four-photon GHZ state:
\begin{equation}\label{GHZ4}
|\Psi^{(4)}_{GHZ}\rangle=\frac{1}{\sqrt{2}}(|h_1h_2h_3h_4\rangle+|v_1v_2v_3v_4\rangle)\,.
\end{equation}
There is a strict requirement for the combined paths to be
indistinguishable in all degrees of freedom, in order for their
amplitudes to interfere. Temporal indistinguishability is
satisfied by generating the photons with a pulsed laser that
defines their generation time, and carefully matching their
relative path lengths. Fortunately, there is no need for sensitive
phase accuracy, but just an overlap of the pulse envelopes.

This scheme is extendable in a straight forward manner. For
example, if a third entangled photon-pair is added in paths 5 and
6, another PBS can fuse it by combining paths 4 and 5 to create a
six-photon GHZ state \cite{Lu07} (see Fig. \ref{Fig1}a). The
addition of each extra pair requires another passage of the pump
beam through a generating crystal, the adjustment of an additional
delay line, an additional projecting PBS and a detection setup
that includes two additional PBS elements and four more
single-photon detectors. Clearly, this approach is non-scalable in
material resources. Another issue is the non-scalability in
temporal resources. Such experiments typically have an
entangled-photon-pair generation probability of one pair every 100
pump pulses and an overall single-photon detection efficiency of
10-25\%. Thus, the more photons an experiment is trying to
produce, the longer the time it takes to accumulate sufficient
data statistics.

In this Letter, we suggest a new approach that solves the problem
of scalability with material resources. Our scheme uses only a
single entangled photon-pair source, a single delay line and a
single projecting PBS element to create entangled states of any
number of photons. A pump pulse is down-converted in a nonlinear
crystal. When a polarization entangled photon pair is generated,
the right photon is directed to a PBS (see Fig. \ref{Fig1}b),
while the left photon enters a delay line. The delay time $\tau$
is chosen such that if a second entangled photon pair is created
by the next pump pulse, the right photon of the second pair meets
with the left photon of the first pair at the projecting PBS.
Post-selecting the events in which one photon exits at each PBS
output port, projects the two entangled pairs onto a four-photon
GHZ state. The left photon of the second pair arrives at the PBS
after travelling through the delay line. The four spatial modes of
previous schemes \cite{Pan01} (1, 2, 3 and 4, as in Eq.
\ref{GHZ4}) are replaced by two spatial modes (1 and 2 after the
projecting PBS, 1' and 2' before it) and three temporal modes (0,
$\tau$, and $2\tau$):
\begin{equation}\label{GHZ4inTime}
|\Psi^{(4)}_{GHZ}\rangle=\frac{1}{\sqrt{2}}(|h_{1'}^0h_2^\tau h_1^\tau h_{2'}^{2\tau}\rangle+|v_{1'}^0v_2^\tau v_1^\tau v_{2'}^{2\tau}\rangle)\,,
\end{equation}
where the lower and upper indices designate the spatial and the
temporal modes, respectively. The first and last photons are
considered before the projecting PBS. It is possible to convert
the mixed spatio-temporal mode partition to only spatial modes by
fast polarization-independent switches.

The most important point to note is the result of the generation
of a third entangled photon pair from the next pump pulse. The
right photon of this third pair is entangled at the PBS with the
delayed left photon of the second pair. All of the six photons
from the three pairs are now in a six-photon GHZ state. There is
no need for any modification between the setups that create the
four- and the six-photon entangled states. It is also clear by
induction that as long as additional consecutive pairs are
created, larger entangled states can be produced.

The use of a PDC source is merely for demonstration purposes, as
our approach can use any photon entangling source, current or
futuristic \cite{Akopian06,Stevenson06,Salter10}. Therefore, our
scheme does neither address the probabilistic nature of PDC
sources, nor other issues that these sources raise, such as the
spectral distinguishability between different photon pairs
\cite{Mosley08}. Nevertheless, our scheme greatly simplifies the
standard approach, enabling the demonstration of entangled photon
states of high photon numbers.

In order to demonstrate our scheme, we created polarization
entangled photon pairs by the nonlinear type-II parametric
down-conversion process \cite{Kwiat95}. A pulsed Ti:Sapphire laser
source with a 76\,MHz repetition rate is frequency doubled to a
wavelength of 390\,nm and an average power of 400\,mW. The laser
beam is corrected for astigmatism and focused on a 2\,mm thick
$\beta$-BaB$_2$O$_4$ (BBO) crystal. Down-converted photons, with a
wavelength of 780\,nm, are spatially filtered by coupling them
into and out of single-mode fibers, and spectrally filtered by
3\,nm wide bandpass filters.

The delay length is chosen such that it fuses pairs that are
created eight pulses apart. The delay time is 105\,ns, which is
longer than the dead time of 50\,ns of the single-photon detectors
(Perkin Elmer SPCM-AQ4C). The delay line is a free-space delay
line (31.6\,m long), built from high reflecting dielectric
mirrors. The total transmittance is higher than 90\% after 10
reflections. The delay line was designed to cancel
distinguishability that results from the different beam
propagation properties of the short and long paths.

\begin{figure}
\centering\includegraphics[angle=0,width=86mm]{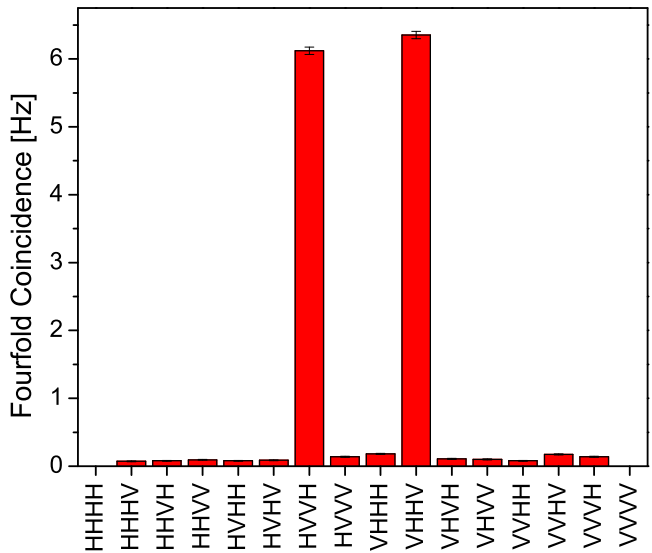}
\caption{\label{Fig2}(Color online) Measured amplitude histogram
of the generated four-photon GHZ state. Two opposite amplitudes
occur on average 65 times more often than any other amplitude.
Data was accumulated over 2000\,sec. We used photon pairs in the
$|\psi^+\rangle$ Bell state. The results are equivalent to using a
$|\phi^+\rangle$ state, up to local polarization rotations.}
\end{figure}

In order to characterize the four-photon state, the four-photon
correlation statistics in the \textbf{HV} (horizontal-vertical
polarization) basis was measured (see Fig. \ref{Fig2}). Two
opposite possibilities (\textbf{HVVH} and \textbf{VHHV}) are much
more probable than the others, as expected from a four-photon GHZ
state. Additionally, measurements in a rotated polarization basis
(plus and minus (\textbf{PM}) $45^\circ$ linear or right and left
(\textbf{RL}) circular) are required in order to demonstrate the
coherence between these two amplitudes. After rotation, there are
16 possible amplitudes, divided into two groups: the even and the
odd amplitude groups, where each polarization appears an even or
an odd number of times, respectively. When the two projected
photons are indistinguishable, one of the amplitude groups
interferes constructively and the other destructively (which one
depends on the specific polarization rotations).

The two projected photons can be rotated individually with wave
plates positioned after the PBS projection. As the first and last
photons are actually measured at the projecting PBS, it would seem
that fast Pockels cells would need to be placed before this PBS in
order to rotate them. Fortunately, there is a way to circumvent
this complication. If the phase of the entangled pairs is tuned to
$90^\circ$ such that their state becomes
\begin{equation}\label{PhiI}
|\phi^{i}\rangle=\frac{1}{\sqrt{2}}(|h_1h_2\rangle+i|v_1v_2\rangle)\,,
\end{equation}
and half-wave plates at $22.5^\circ$ are positioned before the
projecting PBS, the polarization of the first and last photons is
non-locally rotated before the PBS to the circular polarization
basis while that of the two projected photons remains unchanged
(see the Supplemental Material for a detailed calculation for this
rotation \cite{SM}).

\begin{figure}
\centering\includegraphics[angle=0,width=86mm]{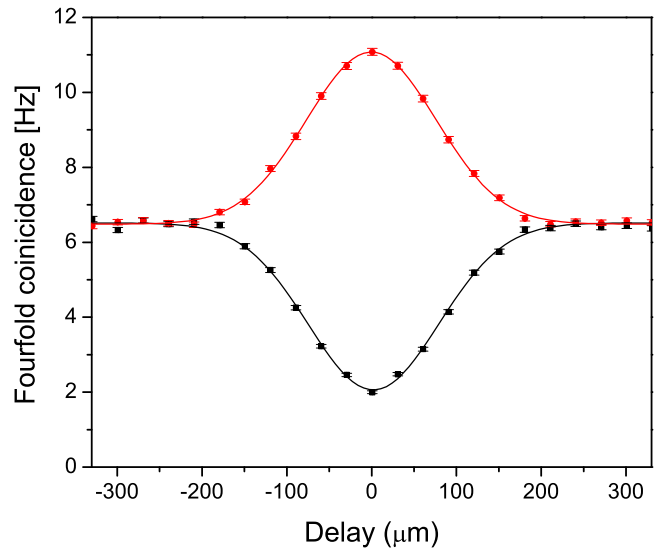}
\caption{\label{Fig3}(Color online) Coherence of the GHZ state.
The sum of all even amplitudes (black squares) and the sum of all
odd amplitudes (red circles) at a rotated polarization basis vs.
the relative delay between the short and the long paths. At zero
delay, the even amplitudes interfere destructively and the odd
amplitudes interfere constructively. When delay is introduced, the
two GHZ amplitudes are temporally distinguishable and interference
is lost. The interference visibility is $69.5\pm0.8$\%. Data was
accumulated over 480\,sec per point.}
\end{figure}

\begin{figure}
\centering\includegraphics[angle=0,width=86mm]{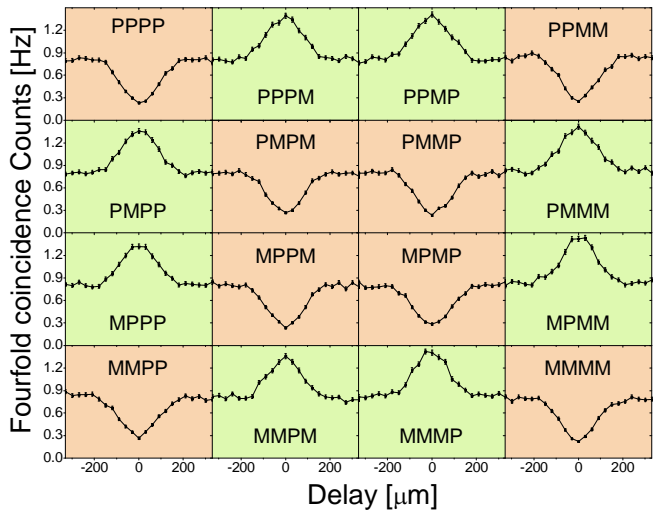}
\caption{\label{Fig4}(Color online) Fourfold coincidence counts of
the 16 individual amplitudes of four photons at the plus-minus
45$^\circ$ (\textbf{PM}) linear polarization basis. At zero delay,
counts from all odd amplitudes (light green background) increase
while counts from all even amplitudes (light orange background)
decrease. Figure \ref{Fig3} is composed of these scans.}
\end{figure}

We applied these rotations and detected all of the 16 even and odd
amplitudes. Each possibility corresponds to different sequences of
the four detectors. Programmable electronics is used to register
the various sequences. The overall post-selected four-photon rate
is 13 events per second. Figure \ref{Fig3} presents the sums of
all counts of the 8 even and of the 8 odd amplitudes (showing
destructive and constructive interference, respectively), as the
delay length is scanned. Figure \ref{Fig4} presents the 16
amplitudes that Fig. \ref{Fig3} is composed of. There are enough
events to observe the interference of all single amplitudes. The
threshold visibility required for the demonstration of
non-locality with four particles is below 35\%
\cite{Mermin90,Zukowski97}. The observed interference visibility
here is $69.5\pm0.8$\%. As the two-photon visibilities are
relatively high (larger than 90\% for HV, PM, and RL measurement
bases at low pump power), the four-photon visibility is an
indication for the projection quality. Nevertheless, the
entangled-pair quality is still a major cause for the fourfold
visibility degradation. We estimate the effect of higher order
terms to cause a degradation of only $\sim3\%$.

The fidelity with a GHZ state can be estimated from the histogram
data of Fig. \ref{Fig2} and the observed fourfold visibility. The
worst case of full coherence between the unwanted diagonal
elements of the state density matrix yields a lower bound of
75.2\% fidelity and assuming no coherence between these terms
results in 79.9\%. As the observed fidelity is higher than 50\%,
it is clear evidence of genuine quantum entanglement between the
generated four photons. We see continuous improvement of the
fourfold visibility and the fidelity as the setup is technically
improved. See the Supplemental Material for a detailed description
of the causes for visibility degradation \cite{SM}.

As the same setup can generate six-photon entanglement, we used it
to detect six-photon states and recorded 30 sixfold events per
hour for a pump power of 500\,mW. As all of the entangled pairs
are produced by the same source, all three pairs have identical
quantum properties. The PBS entangling operations between the
first and second pairs and between the second and third pairs are
identical. Therefore, measuring genuine entanglement between four
photons indicates that entanglement exists between all of the six
detected photons. Quantifying the amount of sixfold entanglement
is left for a future work.

As long as additional pairs are being entangled, a larger GHZ
state is created. Interestingly, when fusing $|\phi^i\rangle$ pair
states (Eq. \ref{PhiI}) instead of $|\phi^+\rangle$, the growing
state is described by a different type of graph from the GHZ graph
\cite{Hein04} (see Fig. \ref{Fig1}c). The four-photon state is
still a GHZ state, but the six-photon state is an H-graph state
\cite{Lu07} (Fig. \ref{Fig1}c), as the left photon from the second
pulse is projected on the PBS after already being rotated to the
PM basis. When yet another (fourth) pair is entangled, the sixth
photon is also rotated, and this pair branches from a corner of
the H-graph .

To conclude, we have demonstrated a new scheme for the generation
of entanglement between many photons. It is efficient in the
required material resources, as the same setup can entangle any
number of photons, without any change. Hence, our scheme can
enable the demonstration of states with larger photon numbers than
what is practically realizable today. Similar to other available
schemes, our scheme suffers from decreasing state production rate
with the number of photons. Nevertheless, there is much room for
improvement as the photon-pair generation probability is currently
between $1-2\%$, but up to $10\%$ is acceptable (see Supplemental
Material \cite{SM}). Photon numbers can be further increased by
using a pump laser with higher power \cite{Yao12}, a generating
crystal with higher nonlinear coefficients \cite{Halevi11},
improving photon collection \cite{Kurtsiefer01} and detection
\cite{Takeuchi99} efficiencies, using detectors with shorter dead
times \cite{Goltsman01,Eraerds07}, lasers with higher repetition
rates \cite{Bartels08} and the coherent addition of pump pulses
\cite{Krischek10}. More virtual qubits can be added to the
generated state by using hyper-entanglement \cite{Kwiat98},
resulting in states of more connected graphs and higher quantum
Hilbert spaces \cite{Gao10}. The incorporation of fast
polarization rotations will enable the generation of different
graph states, as well as on-the-fly alteration of the measurement
basis according to previous measurement outcomes, a procedure
known as \textit{feed-forward} \cite{Prevedel07} which is required
for one-way quantum computation \cite{Raussendorf01}.

The authors thank the Israeli Science Foundation for supporting
this work under grants 366/06 and 546/10.

\end{document}